\documentclass[]{piparticle-final}
\usepackage{graphicx,t1enc}
\usepackage{epsfig}
\usepackage{amssymb,amsmath}

\usepackage{cite} 

\begin{document}

\volume{5}               
\articlenumber{050003}   
\journalyear{2013}       
\editor{G. Mindlin}   
\received{6 April 2013}     
\accepted{3 June 2013}   
\runningauthor{M. N. Kuperman}  
\doi{050003}         

\title{Invited review: Epidemics on social networks}

\author{M. N. Kuperman\cite{inst1,inst2}\thanks{E-mail: kuperman@cab.cnea.gov.ar}
}

\pipabstract{ Since its first formulations almost a century ago, mathematical models for
disease spreading contributed to understand, evaluate and control the epidemic processes.
They promoted a dramatic change in how epidemiologists  thought of the propagation of infectious diseases.
In the last decade, when the traditional epidemiological models seemed to be exhausted,  new types of models were developed.
These new models incorporated concepts from graph theory to describe and model the underlying social structure.
Many of these works  merely produced a more detailed extension of the previous results,  but some others
triggered a completely new paradigm  in the mathematical study of epidemic processes. In this review, we will introduce the basic
concepts of epidemiology, epidemic  modeling and networks, to finally provide a brief description of the most
relevant results in the field. }

\maketitle

\blfootnote{
\begin{theaffiliation}{99}
   \institution{inst1} Consejo Nacional de Investigaciones Cient{\'\i}ficas
y T{\'e}cnicas, Argentina.
   \institution{inst2} Centro At{\'o}mico Bariloche
and Instituto Balseiro, 8400 S. C. de Bariloche, Argentina
\end{theaffiliation}
}

\section {Introduction}

\begin{figure}
\includegraphics[width=\columnwidth]{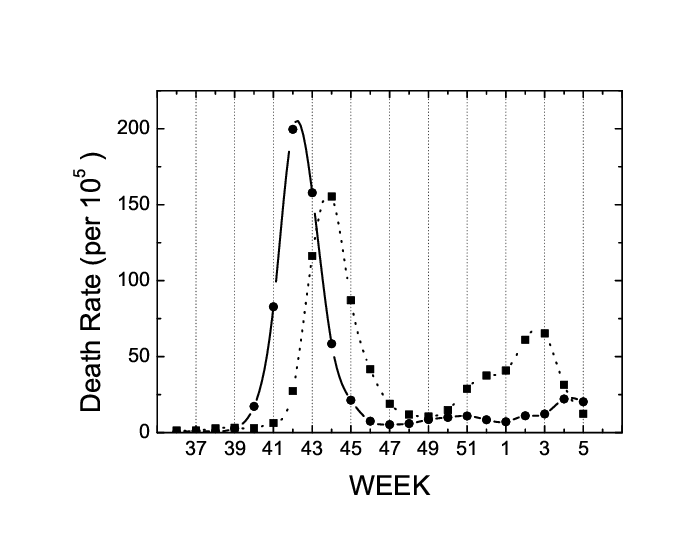}
\caption{Weekly ``Spanish influenza'' death rates in Baltimore
(circles) and San Francisco (squares) from 1918 to 1919. Data
taken from Ref. \cite{brit}.} \label{spflu}
\end{figure}

With the development of more precise and powerful tools, the
mathematical  modeling of infectious diseases has  become a
crucial tool for making decisions associated to policies  on public
health. The scenario was completely different at the beginning of the last
century, when the first mathematical models started to be formulated.
The rather myopic comprehension of the epidemiological processes
was evidenced during the most dramatic epidemiologic events of the last century,
the pandemic 1918 flu. The lack of a mathematical
understanding of the evolution of epidemics gave place to an
inaccurate analysis of the epidemiological situation and subsequent failed assertion
of the success of the immunization strategy. During the influenza pandemic of 1892, a viral disease,
Richard Pfeiffer isolated bacteria from the lungs and sputum of
patients. He  installed, among the medical
community, the idea that these bacteria were the cause of
influenza. At that moment, the bacteria was called Pfeiffer's
bacillus or Bacillus influenzae, while its present name keeps a
reminiscence of Pfeiffer's wrong hypothesis: Haemophilus
influenzae. Though there were some dissenters, the hypothesis of
linking influenza with this pathogen was widely accepted from then
on. Among the supporters of Pfeiffer hypothesis was William Park,
at the New York City Health Department, who in view of the fast progression of the flu in USA,
developed a vaccine and antiserum against Haemophilus influenzae on October 1918.
Shortly afterwards the Philadelphia municipal laboratory released
thousands of doses of the vaccine that was constituted by a  mix
of killed streptococcal, pneumococcal, and H. influenzae bacteria.
Several other attempts to develop similar vaccines followed this
initiative. However, none of these vaccines prevented viral
influenza infection. The present  consensus is that they were even
not protective against the secondary bacterial infections
associated to influenza because the vaccine developers at that
time could not identify, isolate, and produce all the
disease-causing strains of bacteria. Nevertheless, a wrong
evaluation of the evolution of the disease and a lack of
epidemiological knowledge led to the conclusion that the vaccine
was effective. If we look at Fig. \ref{spflu} corresponding to the
weekly influenza death rates in a couple of  U.S. cities taken
from Ref. \cite{brit}, we observe a remarkable decay after vaccination,
in week 43. This decay was inaccurately attributed to the effect
of vaccination as it  corresponds actually to a normal and expected
development of an epidemics without immunization.

The inaccurate association between H. influenzae and influenza
persisted until 1933, when the viral etiology of the flu was
established. But Pfeiffer's influenza bacillus, finally named
Haemophilus influenzae, accounts in its denomination for this
persistent mistake.

The formulation of mathematical models in epidemiology has a
tradition of more than one century. One of the first successful
examples of the mathematical explanation of epidemiological
situations is associated with the study of Malaria. Ronald Ross
was working at the Indian Medical Service during the last years of
the 19th century when he discovered and described the life-cycle of
the malaria parasite in  mosquitoes and developed a mathematical
model to analyze the dynamics of the transmission of the disease
\cite{ross1,ross2,ross3}. His model linked the density of
mosquitoes and the incidence of malaria among the human population. Once
he had identified the anopheles mosquitoes as the vector for
malaria transmission, Ross conjectured that malaria could be
eradicated if the ratio between the number of mosquitoes and the
size of the human population was carried below a threshold value. He
based his analysis on a simple mathematical model.

Ross' model was  based on a set of deterministic coupled
differential equations. He divided the human population into two
groups, the susceptible, with proportion
$S_h$ and the infected, with proportion $I_h$. After recovery,
any formerly infected individual returned to the susceptible
class. This is called a SIS model.  The mosquito population was also
 divided into two groups (with proportions $S_m$ and $I_m$), with no recovery from
infection. Considering equations for the fraction of the
population in each state, we have $S+I=1$ for both humans and
mosquitoes and the model is reduced to a set of two coupled
equations

\begin{eqnarray}
\label{ross}
\frac{dI_h}{dt}&=&a b f I_m (1-I_h)-rI_h\\
 \nonumber
\frac{dI_m}{dt}&=&a c I_h (1-I_m)-\mu_mI_m,
\end{eqnarray}
where $a$ is the man biting rate, $b$ is the proportion of bites
that produce infection in humans, $c$ is the proportion of bites
by which one susceptible mosquito becomes infected, $f$ is the
ratio between the number of female mosquitoes and humans, $r$ is
the average recovery rate of human and $\mu_m$ is the rate of
mosquito mortality.

One of the parameters to quantify the intensity of the epidemics
propagation is the basic reproductive rate $R_0$, that measures
the average number of cases produced by an initial case throughout its
infectious period. $R_0$ depends on several factors. Among them, we can mention
the survival time of an infected individual, the  necessary dose for infection,
the duration of infectiousness in the host, etc. $R_0$ allows to determine whether or not an
infectious disease can spread through a population: an infection
can spread in a population only if $R_0> 1$ and can be maintained
in an endemic state when $R_0 = 1$ \cite{may1}. In the case of
malaria, $R_0$ is defined as the number of secondary cases of
malaria arising from a single case in an susceptible population.
For the model described by Eq. (\ref{ross})

\begin{equation}
R_0=\frac{ma^2bc}{r\mu_m}.
\end{equation}
It is clear that the choice of the parameters affects $R_0$. The
main result is that it is possible to reduce $R_0$ by increasing
the mosquito mortality and reducing the biting rate. For his work on
malaria, Ross was awarded the Nobel Prize in 1902.

Ross' pioneering work was later extended to include other
ingredients and enhance the predictability power of the original
epidemiological model \cite{may1,mcd,aron,diet,aron2,fili,rodri}.

Some years after Ross had proposed his model,  a couple of seminal
works established the basis of the current trends in mathematical
epidemiology. Both models consider the population divided into
three epidemiological groups or  compartments: susceptible (S),
infected (I) and recovered (R).

On the one hand,  Kermack and McKendrick \cite{kerm1} proposed a
SIR model that expanded Ross' set of differential equations. The
model did not  consider the existence of a vector, but a direct
transmission from an infected individual to a susceptible one. A
particular case of the original model,  in which there is no age
dependency of the transmission and recovery rate, is the classical
SIR model that will be explained later.

On the other hand, Reed and Frost \cite{abbe} developed a  SIR
discrete and stochastic epidemic model to describe the
relationship between susceptible, infected and recovered immune
individuals in a population. It is a chain binomial model of
epidemic spread that was intended mainly for teaching  purposes, but that
is the starting point of many modern epidemiological studies. The
model can be mapped into a recurrence equation that defines
what will happen at a given moment depending on what has happened
in the previous  one,

\begin{equation}
I_{t+1} =S_t(1-(1-\rho)^{I_t}),
\end{equation} 
where $I_t$ is the number of cases at time $t$, $S_t$ is the number of susceptible individuals at
time $t$ and $\rho$ is the probability of contagion.

The basic assumption of these SIR models,  which is present in
almost any epidemiological work, is that the infection is spread
directly from infectious individuals to susceptible ones after  a
certain type of interaction between them. In turn, these newly
infected individuals will develop the infection to become
infectious.  After a defined period of time, the infected
individuals heal and remain permanently immune. The interaction
between any two individuals of the population is considered as a
stochastic process with a defined probability of occurrence that
most of the deterministic model  translates  into a contact
rate.

Given a closed population and the number of individuals in each
state, the calculation of the evolution of the epidemics is
straightforward. The epidemic event is over when no infective
individuals remain.

While many classic deterministic epidemiological models were
having success at describing the dynamics of an infectious disease
in a population, it was noted that many involved processes could
be better described by stochastic considerations and thus a new
family of stochastic models  was developed
\cite{bay,ball,andst,diek,ish,tuc}.
 Sometimes, deterministic models introduce some colateral mistakes due
to the continuous character of the involved quantities.An example
of such a case is discussed in Ref. \cite{mollim}. In Ref.
\cite{murr}, the authors proposed a deterministic model to describe
the prevalence of rabies among foxes in England. They predicted a sharp decaying
prevalence of the rabies up to negligible levels, followed by an unexpected new outbreak of infected foxes.
The spontaneous outbreak  after the apparent disappearing of the rabies is due to a fictitious very low endemic
level of infected foxes, as explained in Ref. \cite{mollim}.
The former one is one among several examples of how stochastic models contributed to a
better understanding and explanation of some observed phenomena but,  as their predecessors, they
considered a mean field scheme in the set of differential equations.

Traditional epidemiological models have successfully describe the
generalities of the time evolution of epidemics, the differential
effect on each age group, and some other relevant aspects of an
epidemiological event. But all of them are based on a fully-mixing
approximation, proposing that each individual has the same
probability of getting in touch with any other individual in the
population. The real underlying pattern of social contacts
shows that each individual has a finite set of
acquaintances that serve as channels to promote the contagion.
While the fully mixed approximation allows for writing down a set
of differential equations and a further exploitation of a powerful
analytic set of tools,  a better description of the structure of
the social network provides the models with the capacity  to
compute the epidemic dynamics at the population scale from the
individual-level behavior of infections, with a more accurate
representation of the actual contact pattern. This, in turn,
reflects  some emergent  behavior that  cannot be reproduced
with a system based on a set of differential equation under the
fully mixing assumption. One of the most representative examples
of  this behavior is the so called herd immunity, a form of immunity that occurs when the vaccination of a significant portion of the population
is enough to block the advance of the infection on other non vaccinated individuals.
 Additionally, some network models allow
also for an analytic study of the described process. It is not surprising then that during the
last decade, a new tendency in epidemiological  modeling emerged
together with the inclusion of complex networks as the underlying
social topology in any epidemic event. This new approach proves to
contribute with a further understanding of the dynamics of an
epidemics and  unveils the crucial effect of the social
architecture in the propagation of any infectious disease.

In the following section, we will introduce some generalities
about traditional epidemiological models. In section III, we will
present the most commonly used complex networks when formulating
an epidemiological model. In section IV, we will describe the most
relevant results obtained by  modeling  epidemiological processes
using complex networks to describe the social topology. Next, we will
introduce the concept of herd protection or immunity and a discussion of
some of the works that treat this phenomenon.

\section{Basic Epidemiological Models}

Two main groups can be singled out among the deterministic models for the spread of infectious
diseases which are transmitted through person-to-person contact: the SIR and the SIS.
The names of these models are related to the different groups
considered as components of the population or epidemiological compartments: S corresponds to
susceptible, I to infected and R to removed. The S group
represents the portion of the population that has not been
affected by the disease but may be infected in case of contact
with a sick person. The I group corresponds to those individuals
already infected and who are also responsible for the transmission
of the disease to the susceptible group. The removed group R
includes those individuals recovered from the disease who have temporary or
permanent immunity or, eventually, those  who have died from the
illness and not from other causes. These models may or may not
include the vital dynamics, associated with birth and death
processes. Its inclusion depends on the length of time over which
the spread of the disease is studied.

\subsection{The SIR Model}

As mentioned before, in 1927,  Kermack and McKendrick \cite{kerm1}
developed a mathematical model in which they considered a constant
population divided into three epidemiological groups :
susceptible, infected and recovered. The equations of a SIR model
are

\begin{eqnarray}
\label{sir1} \nonumber \frac{dS}{dt} &=& - \beta S I\\
\frac{dI}{dt} &=& \beta S I - \gamma I\\
\nonumber \frac{dR}{dt} &=& \gamma I,
\end{eqnarray}
where the involved quantities are the proportion of individuals in each group.
As the population is constant,

\begin{equation}
S(t) + I(t) + R(t)=1.
\end{equation}

The SIR model is used when the disease under study confers
permanent immunity to infected individuals after recovery or, in
extreme cases, it kills them. After the contagious period, the
infected individual recovers and is included in the R group. These
models are suitable to describe the  behavior of epidemics
produced by virus agent diseases (measles, chickenpox, mumps, HIV,
poliomyelitis) \cite{heth1}.

The model formulated through Eq. (\ref{sir1}) assumes that all the
individuals in the population have the same probability  of
contracting the disease with a rate of $\beta$, the contact rate.
The number of infected increases proportionally to
both the number of infected and susceptible. The rate of recovery
or removal is proportional to the number of infected only.
$\gamma$  represents the mean recovery rate, ( $1/\gamma$ is the
mean infective period). It is assumed that the incubation time is
negligible and that the rates of infection and recovery are much
faster than the characteristic times associated to births and
deaths. Usually, the initial conditions are set as

\begin{equation}
S(0)>0,\, \, \, I(0)>0\,\, \mbox{and } R(0)=0.
\end{equation}
It is straightforward to show that

\begin{equation}
\left. \frac{dI}{dt}\right|_{t=0} = I(0)(\beta S(0)-\gamma),
\end{equation}
and that the sign of the derivative depends on the value of
$S_c=\frac{\gamma}{\beta}$. When $S(t)>S_c$, the derivative is
positive  and the number of infected individuals increases. When
$S(t)$ goes  below this threshold, the  epidemic starts to fade
out.

A rather non intuitive result can be obtained from  Eq.
\ref{sir1}. We can write

\begin{eqnarray}
\frac{dS}{dR}&=&-\frac{S}{\rho} \nonumber \\
&\Rightarrow& S=S_0 \exp [-R/\rho] \geq S_0 \exp[-N/\rho]>0 \nonumber \\
&\Rightarrow& 0<S(\infty)\leq N. \label{sinfty}
\end{eqnarray}
The epidemics stops when $I(t)=0$, so we can set $I(\infty)=0$, so
$R(\infty)=N-S(\infty)$. From (\ref{sinfty}),

\begin{align}
S(\infty)&=S_0\exp\left[-\frac{R(\infty)}{\rho}\right] \notag\\
&=S_0\exp\left[-\frac{N-S(\infty)}{\rho}\right].
\label{eq:sinfinity}
\end{align}
The last equation is a transcendent expression with a positive
root $S(\infty)$.

Taking (\ref{eq:sinfinity}), we can calculate the total number of
susceptible individuals throughout the whole epidemic process

\begin{equation}
I_{\mbox{\small total}}=I_0+S_0-S(\infty).
\end{equation}

As  $I(t)\to 0$ and $S(t)\to S(\infty)>0$, we conclude that when
the epidemics  end, there is a portion of the population that has
not been affected

The previous model can be extended to include vital dynamics
\cite{kup1},  delays equations \cite{bere},  age structured
population, migration \cite{franc}, and diffusion. In any case,
all these generalizations only introduce some slight changes on
the steady states of the system, or in the case of spatially extended  models,
travelling waves \cite{yan}.

\begin{figure}
\includegraphics[width=7.cm]{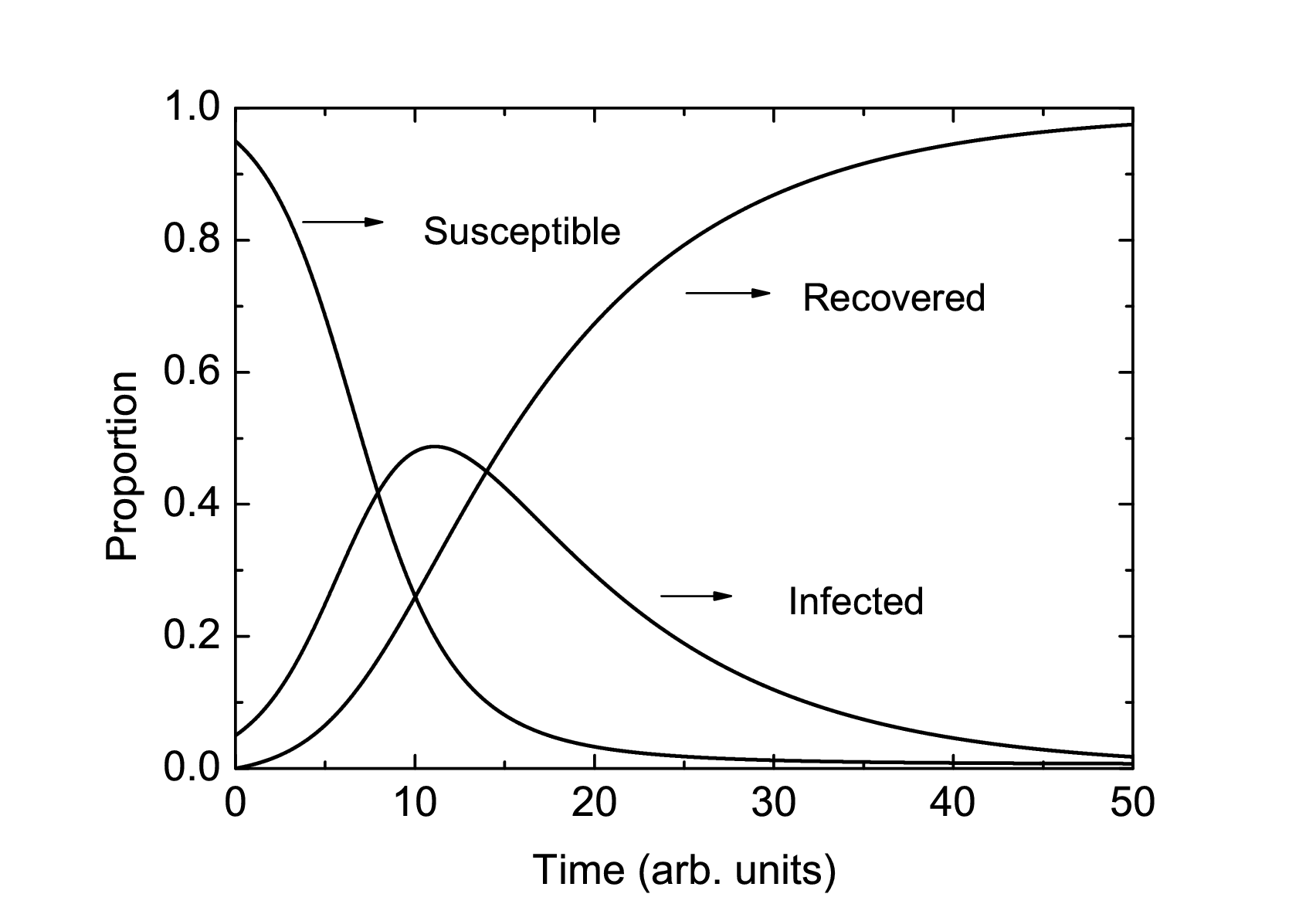}
\caption{Temporal  behavior of the proportion of individuals in
each of the three compartments of the SIR model.} \label{sisg}
\end{figure}

Figure \ref{sisg}  displays  the typical  behavior of the density of
individuals in each of the epidemiological compartments described
by Eq. (\ref{sir1}). Compare this with the pattern shown in Fig.
\ref{spflu}.

\subsection{The SIS Model}

The SIS model assumes that the disease does not confer immunity to
infected individuals after recovery. Thus, after the infective
period, the infected individual recovers and is again included in
the S group. Therefore, the model presents only two
epidemiological compartments, S and I. This model is suitable to
describe the  behavior of epidemics produced by bacterial agent
diseases (meningitis, plague, venereal diseases) and by protozoan
agent diseases (malaria) \cite{heth1}. We can write the equations
for a general SIS model assuming again  that the population is
constant,

\begin{eqnarray}
\label{sis1} \nonumber \frac{dS}{dt} &=& - \beta S I + \gamma I \\
\frac{dI}{dt} &=& \beta S I - \gamma I.
\end{eqnarray}
As the relation $S + I =1$ holds, Eq. (\ref{sis1}) can be reduced to
a single equation,

\begin{equation}
\label{sis2} \frac{dI}{dt} = (\beta-\gamma)I-\beta I^2.
\end{equation}
The solution of this equation is

\begin{equation}
\label{sis3} I(t) = (1-\frac{\gamma}{\beta}) \frac{C
\exp[(\gamma-\beta)t]}{1+C\exp[(\gamma-\beta)t]},
\end{equation}
where $C$ is defined by the initial conditions as

\begin{equation}
\label{sis4} C = \frac{\beta i_0}{\beta(1-i_0)-\gamma}.
\end{equation}

If $I_0$ is small and $\beta>\gamma$, the solution is a logistic
growth that saturates before the whole population is infected, the
stationary value is $I_s=\frac{\beta-\gamma}{\beta}$. It can be
shown that $R_0=\beta/\gamma$. This sets the condition for the
epidemic to persist.

\subsection{Other models}
The literature on epidemiological models includes several
 generalizations  about the previous ones to adapt the description to
the particularities of a specific infectious disease \cite{heth2}.
One possibility is to increase the number of compartments to
describe different stages of the state of an individual during the
epidemic spread. Among these models, we can mention the SIRS, a
simple extension of the SIR that does not confer a permanent
immunity to recovered individuals and after some time they rejoin
the susceptible group,

\begin{eqnarray}
\label{sirs} \nonumber \frac{dS}{dt} &=& - \beta SI + + \lambda R\\
\frac{dI}{dt} &=& \beta SI - \gamma I\\
\nonumber \frac{dR}{dt} &=& \gamma I - \lambda R.
\end{eqnarray}

Other models include more epidemiological groups or compartments,
such as the SEIS and SEIR model, that take into consideration the
exposed or latent period of the disease, by defining an additional
compartment E.

There are several diseases  in which there is a vertical transient
immunity transmission from a mother to her newborn. Then, each
individual is born with a passive immunity acquired from the
mother. To indicate this, an additional group P is added.

\begin{figure}
\includegraphics[width=\columnwidth]{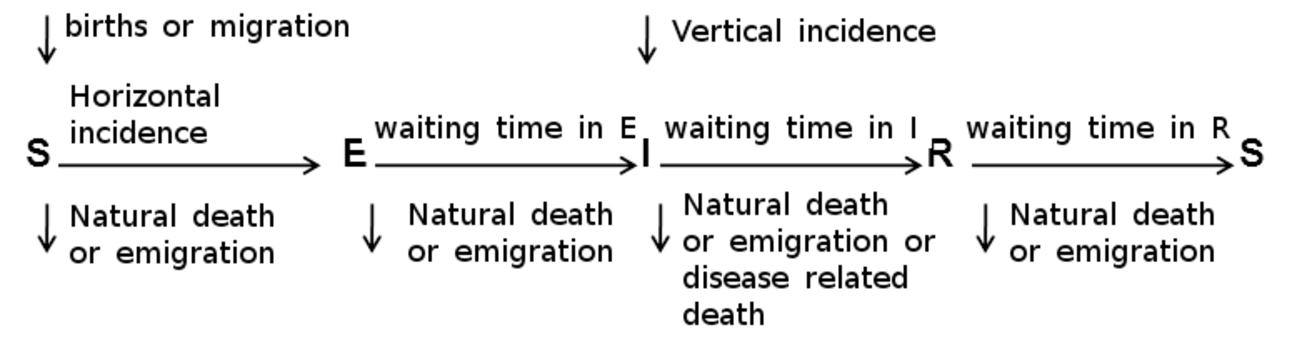}
\caption{Transfer diagram for a SEIRS models. Taken from
Ref. \cite{heth2}. } \label{flow}
\end{figure}

The range of possibilities is rather extended, and this is
reflected in the title of Ref. \cite{heth2}: ``A thousand and one
epidemiological models''. There are a lot of possibilities to
define the compartment structure. Usually, this structure is
represented as a transfer chart indicating the flow between the
compartments and the external contributions. Figure \ref{flow} shows
an example of a diagram for a SEIRS model, taken from
Ref. \cite{heth2}.

Horizontal incidence refers to a contagion due to a contact
between a susceptible and infectious individual, vertical
incidence account for the possibility for the offspring of
infected parents to be born infected, such as with AIDS, hepatitis
B, Chlamydia, etc.

Many of the previous models have been expanded, including
stochastic terms. One of the most relevant differences between the
deterministic and stochastic models is their asymptotic  behavior.
A stochastic model can show a solution converging to the
disease-free state when the deterministic counterpart predicts an
endemic equilibrium. The results obtained from the stochastic
models are generally expressed  in terms of the probability of an
outbreak and  of its  size and duration distribution
\cite{bay,ball,andst,diek,ish,tuc}.

\section{Complex Networks}

A graph or network is a mathematical representation of a set of
objects that may be connected between them through links.  The
interconnected objects are represented by the nodes (or vertices)
of the graph while the connecting links are associated to the
edges of the graph. Networks can be characterized by several
topological properties, some of which will be introduced later.
Social links are preponderantly non directional (symmetric), though there are
some cases of social directed networks. The set of nodes
attached to a given node through these links is called  its
neighborhood. The size of the neighborhood is the degree of the
node.

While the study of graph theory dates back to the pioneering works
of Erdös and Renyi in the  1950s \cite{erd}, their gradual
colonization of the modern epidemiological models has only started a
decade ago. The attention of modelers was drawn to graph theory
when some authors started to point out that the social structure
could be mimicked by networks constructed under very simple
premises \cite{ws,ba}.  Since then, a huge collection of computer-generated
networks have been studied in the context of disease transmission.
The underlying rationale for the use of networks is that they can
represent how individuals are distributed in social and
geographical space and how the contacts between them are promoted,
reinforced  or inhibited, according to the rules of social
dynamics. When the population is fully mixed, each individual has
the same probability of coming into contact with any other
individual. This assumption makes it possible to calculate the
effective contact rates $\beta$ as the product of the transmission
rate of the disease, the effective number of contacts per unit
time and the proportion of these contacts that propagate the
infection. The formulation of a mean field model is then
straightforward. However, in real systems, the acquaintances of
each individual are reduced to a portion of the whole population.
Each person has a set of contacts that shapes the local
topology of the  neighborhood. The whole social architecture, the
network of contacts,  can be represented with a graph.

In the limiting case when the mean degree of the nodes in a network is
close to the total number of nodes, the difference between a structured
population and  a fully mixed one fades out. The differences are
noticeable when the network is diluted, i.e., the mean degree of
the node is small compared with the size of the network. This will
be a necessary condition for all the networks used to model
disease propagation. In the following paragraphs, we will introduce
the most common families of networks used  for epidemiological
 modeling.

\begin{figure}[!h]
\centerline{\includegraphics[clip=true,scale=0.25]{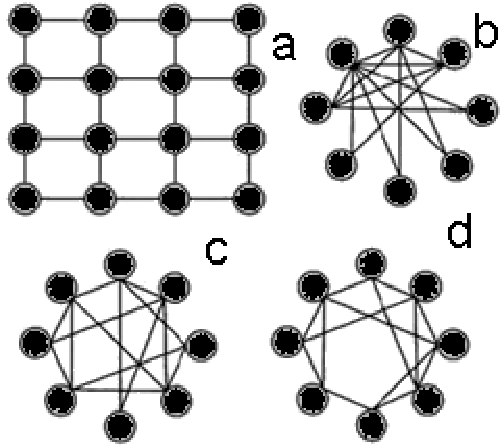}}
\caption{Scheme of  four kinds of networks: (a) Lattice, (b)scale
free, (c) Exponential, (d) Small World.} \label{fn4}
\end{figure}

{\bf Lattices.} When incorporating a network to a model, the
simplest case is considering a grid or a lattice. In a squared $d$
dimensional lattice, each node is connected to  $2d$  neighbors.
Individuals are regularly located and connected with adjacent
 neighbors; therefore, contacts are localized in space. Figure
\ref{fn4} shows, among others, an example of a two dimensional
square lattice

\begin{figure}[!h]
\centerline{\includegraphics[clip=true,scale=0.16]{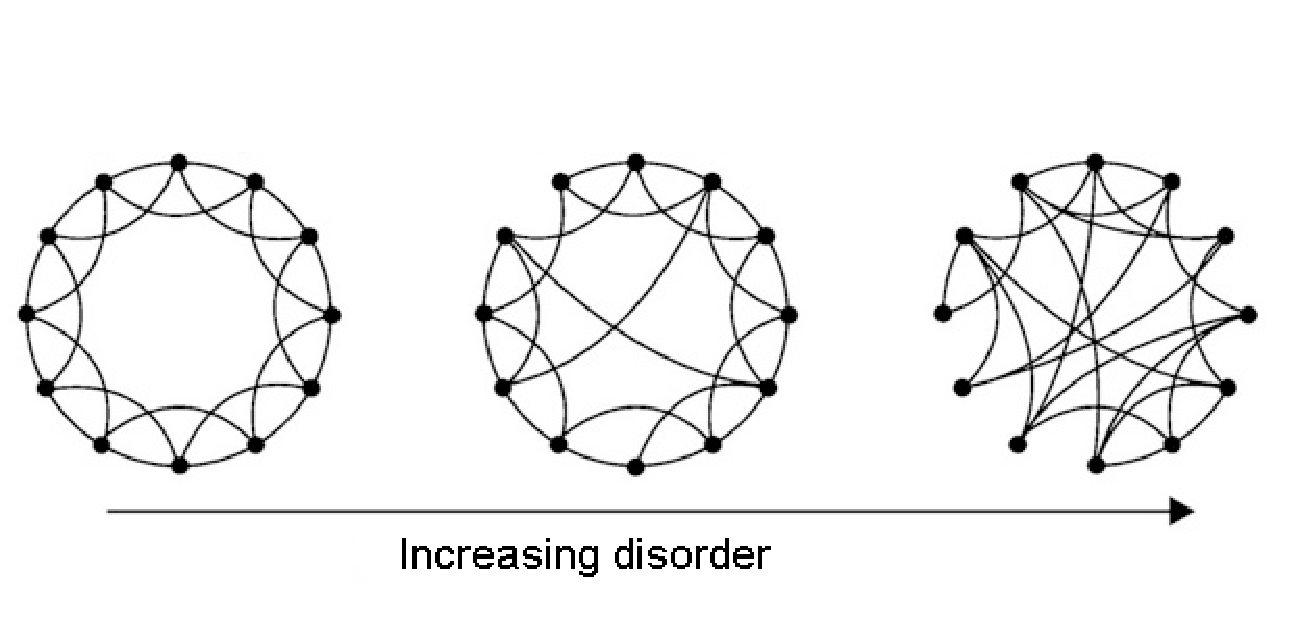}}
\caption{Representation of several Small World Networks
constructed according the algorithm presented in Ref. \cite{ws}. As the
disorder degree increases, there number of shortcuts  grow
replacing some of the original (ordered network) links.} \label{ws}
\end{figure}

{\bf Small-world networks.} The concept of {\it Small World} was
introduced by Milgram in 1967 in order to describe the topological
properties of social communities and relationships \cite{milg}.
Some years ago, Watts and Strogatz introduced a model for
constructing networks displaying topological features that mimic
the social architecture revealed by Milgram. In this model of
Small World (SW) networks a single parameter $p$, running from 0
to 1, characterizes the degree of disorder of the network, ranging
from a regular lattice to a completely random graph \cite{ws}. The
construction of these networks starts from a regular,
one-dimensional, periodic lattice of N elements and coordination
number $2K$. Each of the sites is visited, rewiring K of its links
with probability $p$. Values of $p$ within the interval [0,1]
produce a continuous spectrum of small world networks. Note that
$p$ is the fraction of modified regular links. A schematic
representation of this family of networks is shown in Fig.
\ref{ws}.

To characterize the topological properties of the SW networks, two
magnitudes are calculated. The first one, $L(p)$, measures the
mean topological distance between any pair of elements in the
network, that is, the shortest path between two vertices, averaged
over all pairs of vertices. Thus, an ordered lattice has $L(0)\sim
N/K$, while, for a random network, $L(1)\sim ln(N)/ln(K)$. The
second one, $C(p)$, measures the mean clustering of an element's
neighborhood. $C(p)$ is defined in the following way: Let us
consider the element $i$, having $k_i$ neighbors connected to it.
We denote by $c_i(p)$ the number of neighbors of element $i$ that
are neighbors among themselves, normalized to the value that this
would have if all of them were connected to one another; namely,
$k_i(k_i-1)/2$. Now, $C(p)$ is the average, over the system, of the
local clusterization $c_i(p)$. Ordered lattices are highly
clustered, with $C(0)\sim 3/4$, and random lattices are
characterized by $C(1) \sim K/N$. Between these extremes, small
worlds are characterized by a short length between elements, like
random networks, and high clusterization, like ordered ones.

\begin{figure}[!h]
\centerline{\includegraphics[width=7.cm]{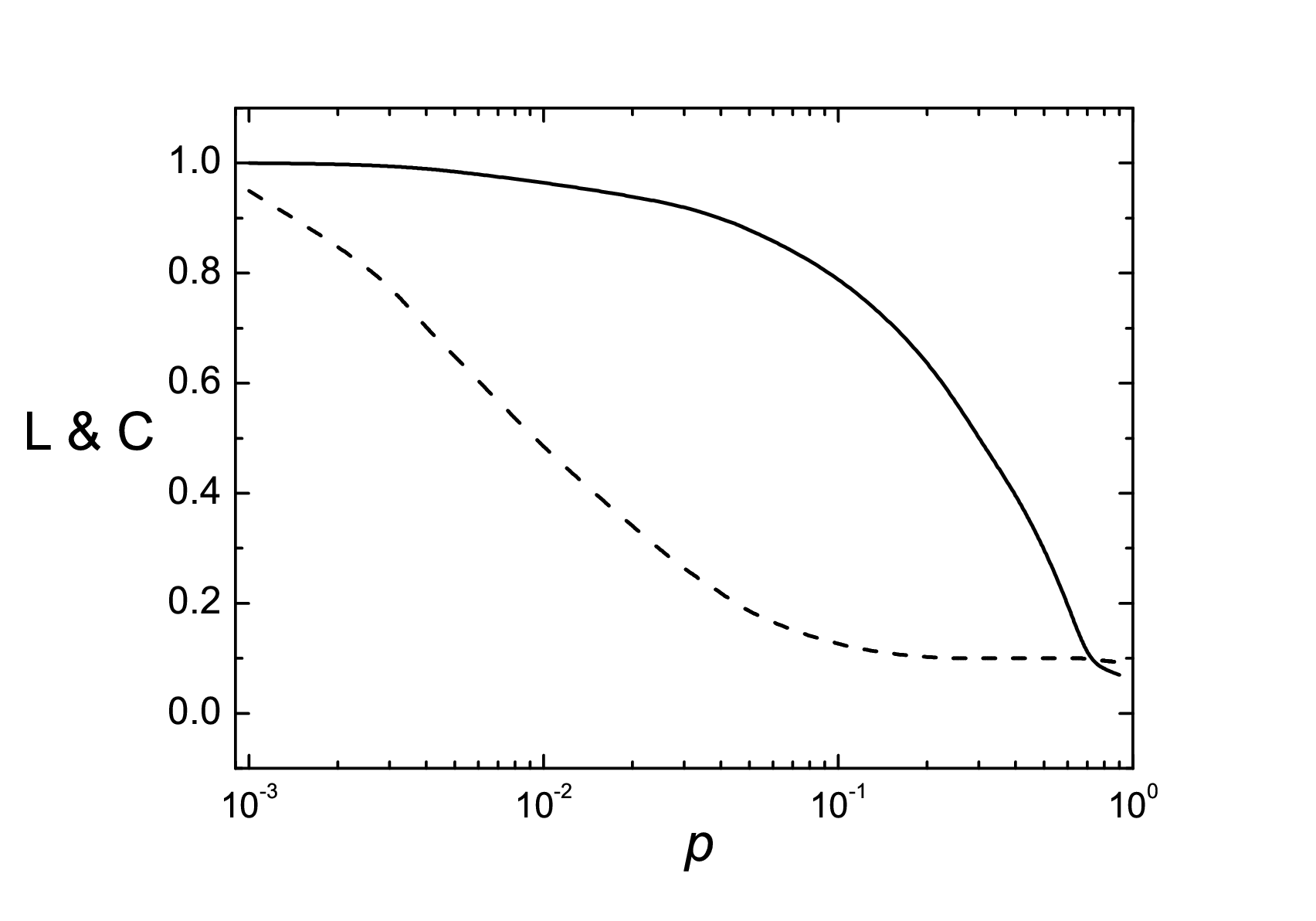}} \caption{In
this figure, we show the mean values of the clustering coefficient
$C$ and the path length $L$ as a function of the disorder
parameter $p$. Note the fast decay of $L$ and the presence of a
region where the value adopted by $L$ is similar to the one
corresponding to total disorder, while the value adopted by $C$ is
close to the one corresponding to the ordered case.} \label{swp}
\end{figure}

\begin{figure}[!h]
\centerline{\includegraphics[width=7.cm]{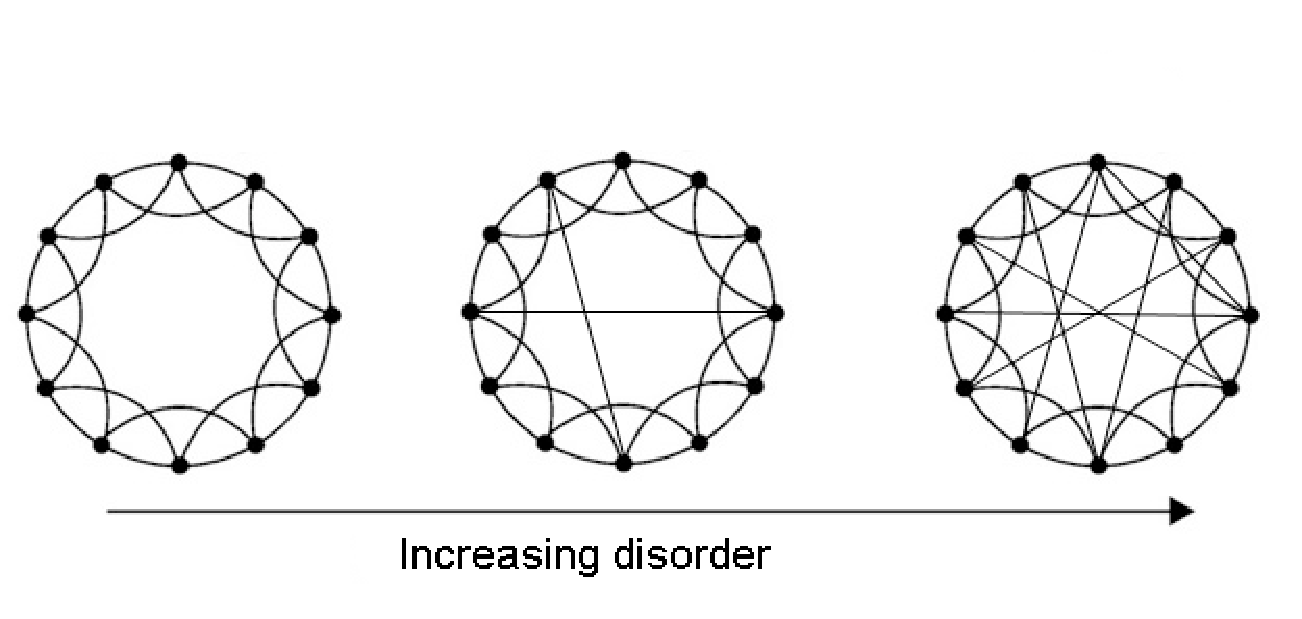}}
\caption{Representation of several Small World Networks
constructed according the algorithm presented in Ref. \cite{nw}. As the
disorder degree increases,  three number of shortcuts as well as
the number of total links grow. } \label{nw}
\end{figure}

Other procedures for developing similar social networks have been
proposed in Ref. \cite{nw} where instead of rewiring existing links to
create shortcuts, the procedure add links connecting two randomly
chosen nodes with probability $p$. In Fig. \ref{nw}, we show an
example, analogous to the one shown in Fig. \ref{ws}.

{\bf Random networks.} There are  different families of networks
with random genesis but displaying a wide spectra of complex
topologies. In random networks, the spatial position of
individuals is irrelevant and the links are randomly distributed.
The iconic Erd\"os-R\'enyi (ER) random graphs are built from a set
of nodes that are randomly connected with probability $p$,
independently of any other existing connection. The degree
distribution, i.e., the number of links associated to each node, is
binomial and  when the number of  nodes is large, it can be approximated by a Poisson distribution
\cite{newmanb}.
\begin{figure}[!h]
\centerline{\includegraphics[width=\columnwidth]{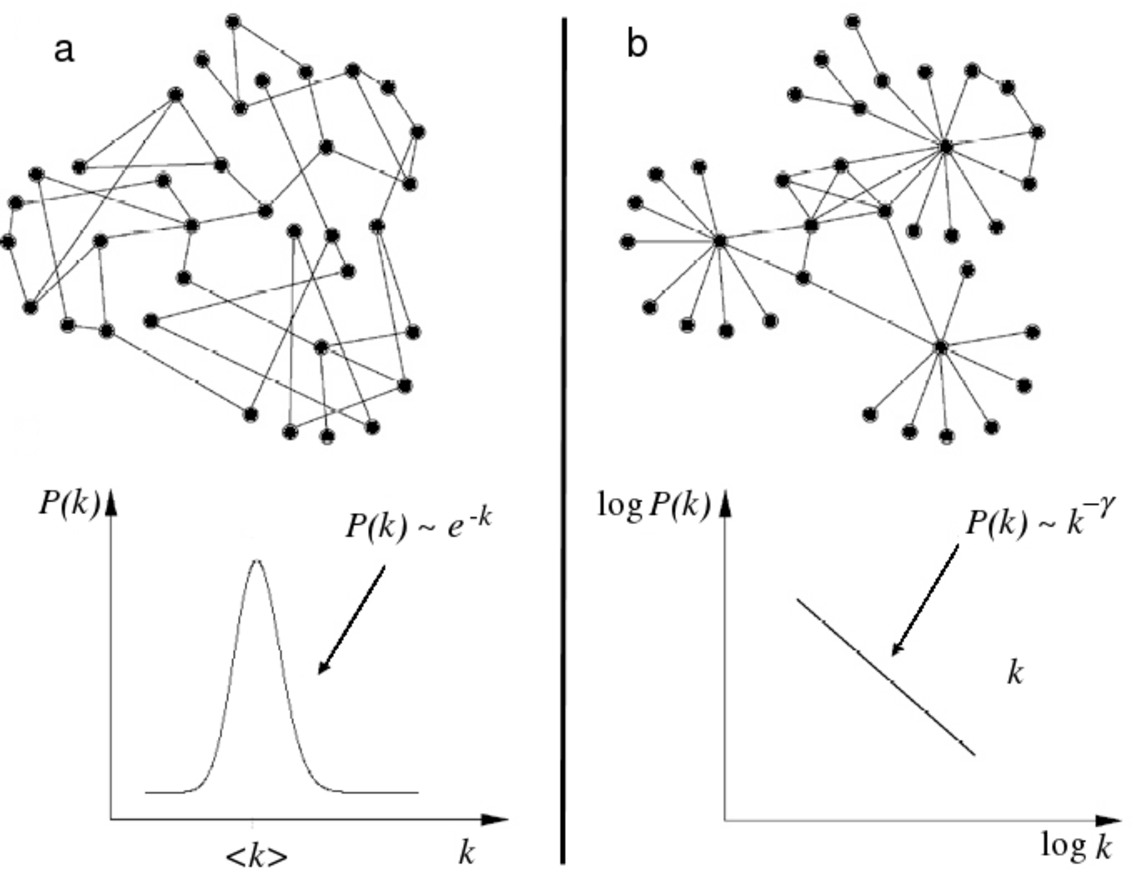}} \caption{This
figure shows examples of (a) ER and (b) BA networks. The figure also
displays the connectivity distribution $P(k)$, that follows a
binomial distribution for the ER networks and a power law for BA
networks.} \label{rnet}
\end{figure}
In Ref. \cite{gener}, the authors propose a formalism based on the
generating function that permits to construct random networks with
arbitrary degree distribution. The mechanism of construction
also allows for further analytic studies on these networks. In
particular, networks can be chosen to have a power law degree
distribution. This case will be presented in the next paragraphs.

{\bf Scale-free network.} As mentioned before, one of the most
revealing measures of a network is its degree of distribution, i.e.,
the distribution of the number of connections of the nodes. In
most real networks, it is far from being homogeneous, with
highly connected individuals on one extreme and almost isolated
nodes  on the other. Scale-free networks provide a means of
achieving such extreme levels of heterogeneity.

Scale-free networks are constructed by adding new individuals to a core,  with
a connection mechanism that imitates the underlying process that rules the choice
of social contacts. The Barab\'asi - Albert (BA) model algorithm, one of the triggers of the present huge interest on
scale-free networks, uses a preferential attachment mechanism \cite{ba}. The algorithm starts from a small
nucleus of connected nodes. At each step, a new node is added to
the network and connected to $m$ existing nodes. The probability
of choosing a node $p_i$ is proportional to the number of links
that the existing node already   has $$p_i = \frac{k_i}{\sum_j
k_j},$$ where $k_i$ is the degree of node $i$. That means that the
new nodes have a preference to attach themselves to the most
``popular'' nodes. One  salient  feature of these networks is
that their degree distribution is scale-free, following a power
law of the form

$$P\left(k\right)\sim k^{-3}.$$

A sketch of the typical topology of the last two networks is shown
in Fig. \ref{rnet}. While  the degree distribution of the ER
network has a clear peak and is close to homogeneous, the topology
of the BA network  is dominated by the presence of hub, highly
connected nodes. The figure also displays the typical degree
distribution $P(k)$ for each case.

 Over the last years, many other attachment  mechanisms have been
proposed to obtain scale-free networks with other adjusted
properties such as the clustering coefficient, higher moments of
the degree distribution \cite{kapri,klem,holm,brun}.

{\bf Coevolutive or adaptive topology.} When one of the former
examples of networks is chosen as a model for the social woven,
there is an implicit assumption:  the underlying social topology is
frozen. However, this situation does not reflect the observed fact
that in real populations, social and migratory phenomena, sanitary
isolation or other processes can lead to a dynamic configuration
of contacts, with some links being eliminated, other being
created. If the time span of the epidemics is long enough, the
social network will change and these changes will not be reflected
if the topology remains fixed. This is particularly important in
small groups. The social dynamics, including the epidemic process,
can  shape the topology of the network, creating a feedback mechanism
that can favor or attempt against the propagation of an infectious
disease. For this reason, some models consider a coevolving
network, with dynamic links that change the aspect of the networks
while the epidemics  occur.

\section{Epidemiological Models on Networks}
In this section, we will discuss several models based on the use of complex networks
to mimic the social architecture. The discussion will be organized according to the topology of these
underlying networks.

\begin{figure}[!h]
\centerline{\includegraphics[width=\columnwidth]{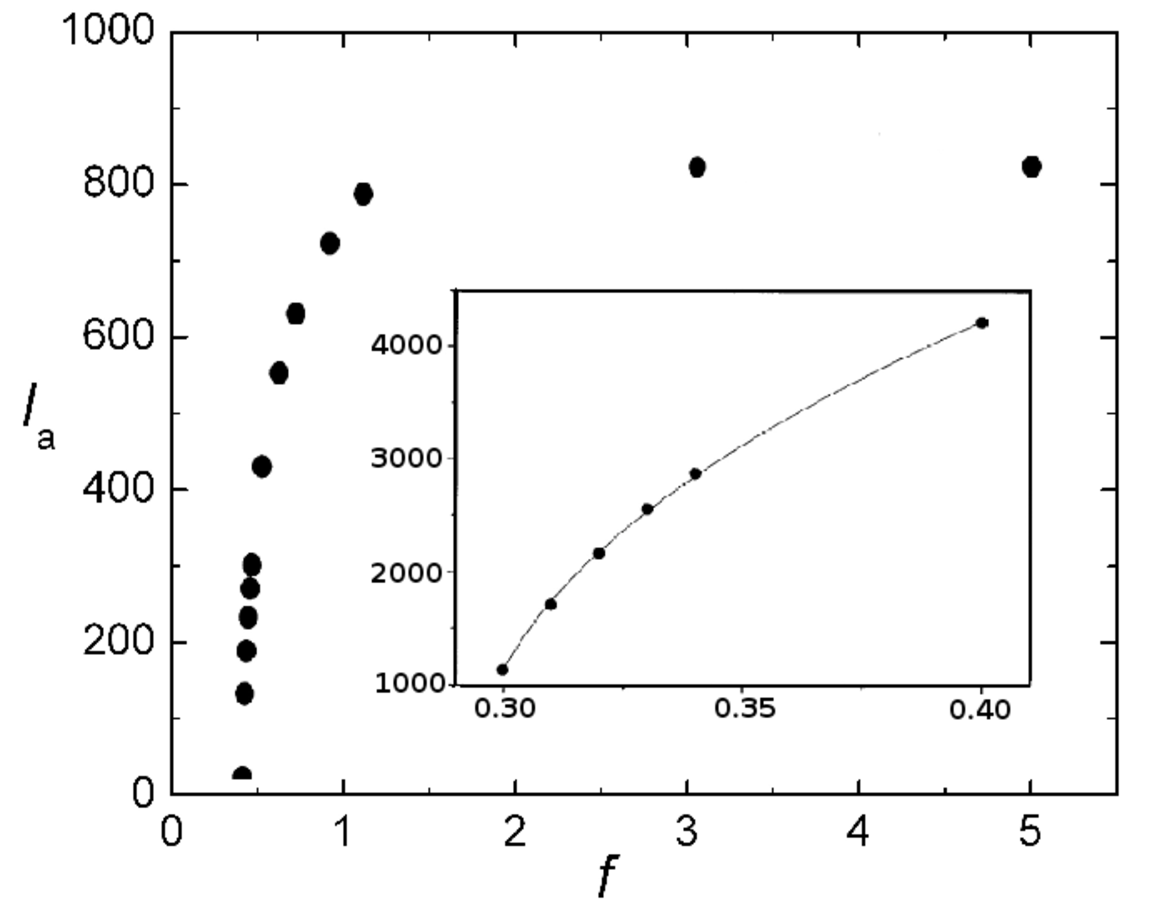}} \caption{ SIS
model. Asymptotic value of infected individuals as a function of
the infectivity of infectious individuals. The inset displays the
scaling of the data with a power-like curve $A|f-f_c|^\alpha$,
with $\alpha \approx 0.5$. Adapted from Ref. \cite{kup2}.}
\label{aut1}
\end{figure}

{\bf Lattices.}  Lattices were the first attempt to represent the
underlying topology of the social contacts and thus to analyze the
possible effect of interactions at the individual level. These
models took distance from the paradigmatic fully mixed assumption
and  focused  on looking for those phenomena that a mean field
model could not explain. Still, the lattices  cannot fully capture
the role of inhomogeneities.  As the individuals are located on a
regular grid, mostly two dimensional,  the neighborhood of each
node is reduced to the adjacent nodes, inducing only short range
or localized interactions.  A typical model considers that the nodes can be in any of the
epidemiological states or compartments. The dynamic of the
epidemics evolves through a contact process \cite{harri} and the
evolutive rules  do not differ too much from traditional cellular
automata models \cite{wolf}. Disease transmission is  modeled as a
stochastic process. Each infected node has a probability $p_i$ of
infecting a  neighboring susceptible node. Once infected, the
individuals may recover from infection with a probability $p_r$;
i.e., the infective stage  lasts typically  $1/p_r$. From the
infective phase, the individuals can move back to the susceptible
compartment or the recovered phase, depending on whether the
models  are SIS or SIR. Usually, a localized infectious focus is
introduced among the population. The transient shows a local and
slow development of the disease that at the initial stage
involves the growing of a cluster, with the infection propagating
at its boundary, like a  traveling wave. After the initial
transient, SIS, SIR and SIRS models behave in different ways. The
initially local dynamics that can or cannot propagate to the whole
system is what introduces a completely new  behavior in this
spatially extended model. In Ref. \cite{grass1}, the author argued the
infective clusters behave as the clusters in the directed
percolation model. Figure \ref{aut1} shows an example of the  behavior of the
asymptotic value of infected individuals under SIS dynamics in a two dimensional
square lattice. The figure reflects the results found in
Ref. \cite{kup2}. The parameter $f$ is associated to the infectivity of
infectious individuals, closely related to the contact rate. We
observe the inset displaying  the scaling of the data with a
power-like curve $A|f-f_c|^\alpha$, with $\alpha \approx 0.5$
\cite{kup2}.

As mentioned before, Kermack and Mckendrick \cite{kerm1} proved
the existence of a propagation threshold for the disease invading
a susceptible population. The lattice based SIR models introduce a
different threshold. The simulations show that epidemics can just
 remain localized around the initial focus or turn into a
pandemic, affecting the entire population. The most dramatic
examples of real pandemic are the Black Plague between the 1300
and 1500 and the Spanish Flu, in 1917-1918. Both left a wake of
death and terror while crossing the European continent. The predicted new
threshold  established a limit  below in which the pandemic behaviour is
not achieved.
\begin{figure}[!h]
\centerline{\includegraphics[width=\columnwidth]{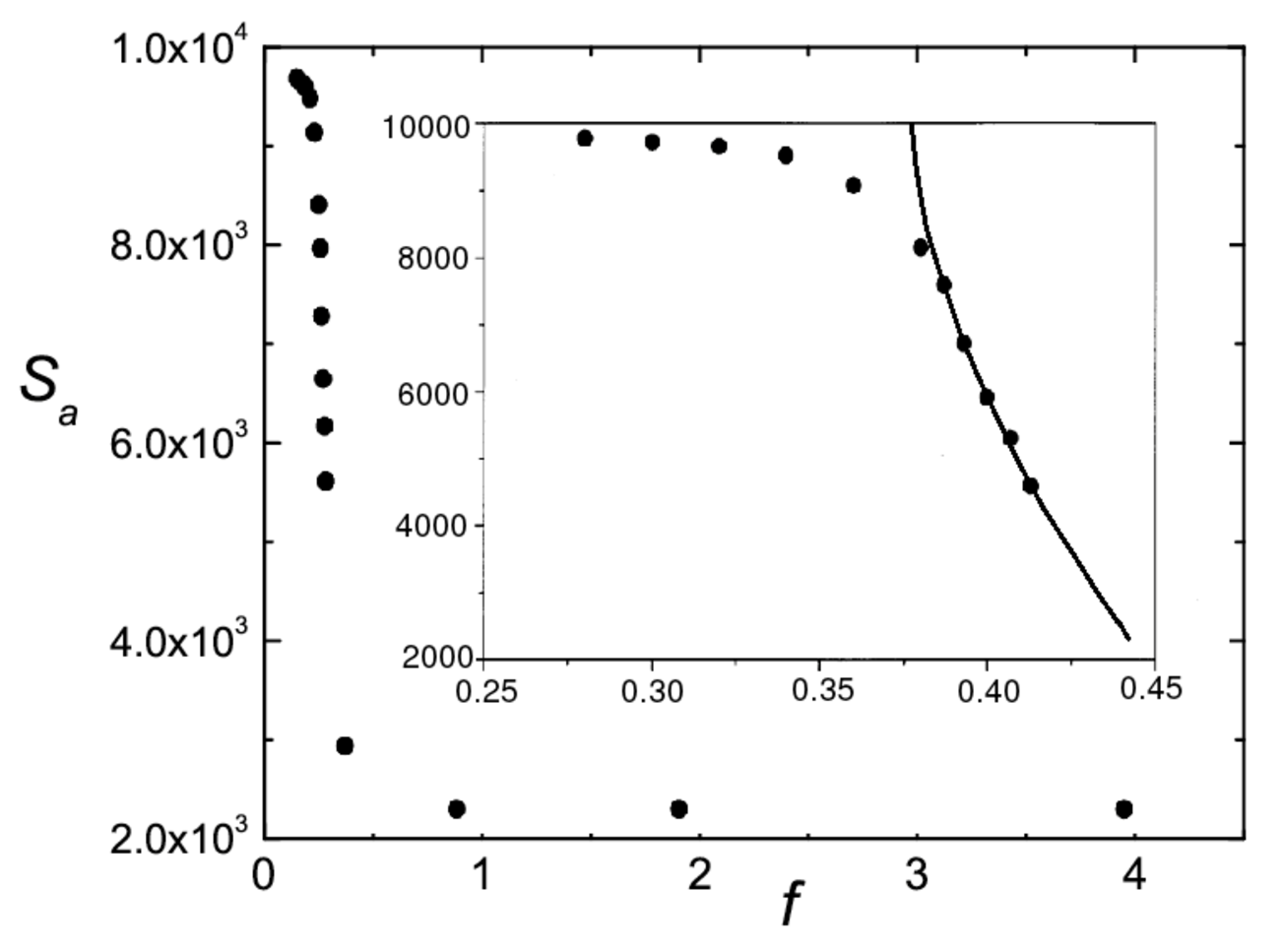}} \caption{ SIR
model. Asymptotic value of susceptible individuals as a function
of the infectivity of infectious individuals. The inset displays
the scaling of the data with a power-like curve $A|f-f_c|^\alpha$,
with $\alpha \approx 0.5$. Adapted from Ref. \cite{kup2}.}
\label{aut2}
\end{figure}

Some works about epidemic propagation on lattices are analogous  to
forest fire models \cite{bak}, with the characteristic feature
that the frequency distributions of the epidemic sizes and
duration obey a power-law. In Ref. \cite{rod1,rod2}, the authors exploit
 these analogies to explain the observed  behavior of measles,
whooping cough and mumps in the Faroe Islands. The observed data
display a power-like  behavior.

{\bf Random networks.} Most of the models based on random graphs
were previous to the renewed interest on complex networks. A
simple but effective idea for the study of the dynamics of
diseases on random networks is the contact process proposed in
Ref. \cite{diek2} that produces a branching phenomena while the
infection propagates. In Ref. \cite{barb}, the authors use a E-R network
with an approximately Poisson degree distribution. A common
feature to all these models is that the rate of the initial
transient growth is smaller than the corresponding to similar
models in  fully-mixed populations. This effect can be easily
understood noting that, on the one hand, the degree of a given
initially infected node is typically small, thus having a limited
number of susceptible contacts. On the other hand, there is a
self limiting process due to the fact that the same infection
propagation predates the local availability of susceptible
targets.

A different analytical approach to random networks is presented in
Ref. \cite{new3}. The author shows that a family of variants of the SIR
model  can be solved exactly on random networks built by a
generating function method and appealing to the formalism of
percolation models. The author analyzes the propagation of a
disease in networks with arbitrary degree distributions and
heterogeneous infectiveness times and transmission probabilities.
The results include the particular case of scale-free networks,
that will be discussed later.

{\bf Small-world networks.}  As mentioned above, regular networks
can exhibit high clustering but long path lengths. On the other
extreme, random networks have a lot of shortcuts between two
distant individuals, but a negligible clustering. Both features
affect the propagative  behavior on any  modeled disease. The
spread of infectious diseases on SW networks has been analyzed in
several works. The interested was triggered  by the fact that even
a small number of random connections added to a regular lattice,
following for example the algorithm described in Ref. \cite{ws},
produces unexpected macroscopic effects. By sharing topological
properties from random and ordered  networks, SW networks can display
complex propagative patterns. On the one hand, the high level of
clustering means that most infection occurs locally. On the other
hand, shortcuts are vehicles for the fast spread of the epidemic
to the entire population.

\begin{figure}[!h]
\centerline{\includegraphics[width=\columnwidth]{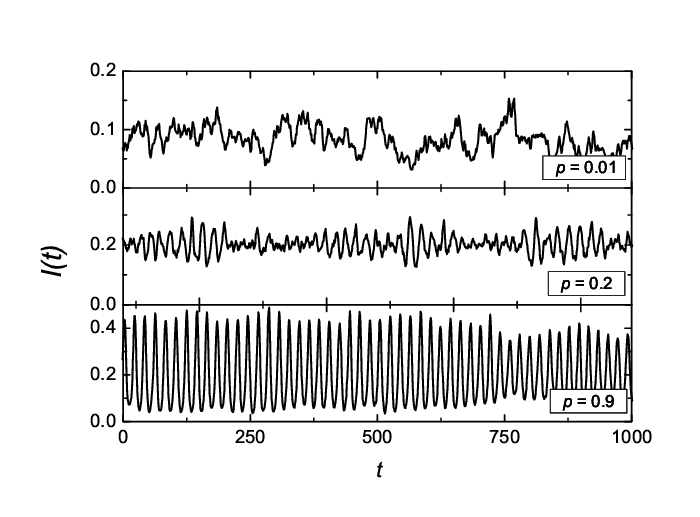}} \caption{
Asymptotic  behavior of the number of infected individuals in
three SW networks with different degrees of disorder $p$. The
emergence of a synchronized pattern is evident in the bottom graph.
} \label{swepi}
\end{figure}

In Ref. \cite{moore}, the authors study a SI model and show that
shortcuts can dramatically increase the possibility of an epidemic
event. The analysis is based on bond percolation concepts. While
the result could be easily anticipated due to the long range
propagative properties of shortcuts, the authors find an important
analytic result. It was a study of a SIRS models that showed for
the first time the evidence of a dramatic change in the  behavior of
an  epidemic due to changes in the underlying social topology
\cite{kup3}. By specifically analyzing the effect of clustering on
the dynamics of an epidemics, the authors show that a SIRS model
on a SW network presents two distinct  types of behavior. As the rewiring
parameter $p$ increases, the system transits from an endemic state,
with a low level of infection to  periodic oscillations in the
number of infected individuals, reflecting an underlying
synchronization phenomena. The transition from one regime to the
other is sharp and occurs at a finite value of $p$. The reason
behind this phenomenon is still unknown. Figure \ref{swepi} shows the
temporal  behavior of the number of infected individuals for three
values of the rewiring parameter $p$, as found in Ref. \cite{kup3}.

It would not be responsible to affirm that SW networks reflect all
the real social structures. However, they capture essential aspects of
such organization that play central roles in the propagation of a
diseases, namely, the clustering coefficient and the short social
distance between individuals. Understanding that there are
certain limitations, SW networks help to mimic different social
organizations that range from rural population to big cities. There
are more sophisticated models of networks with topologies that are
more closely related to real social organizations at large scale.
These networks are characterized by a truncated power law
distribution of the degree of the nodes and by values of
clustering and mean distance corresponding to the small world
regime.

{\bf Scale-free networks.} Scale-free networks captured the attention of  epidemiologists
due to the close resemblance between their extreme degree
distribution and the pattern of social contacts in real
populations. A power law degree distribution presents individuals
with many contacts and who play the role of super-spreaders. A
higher number of contacts implies  a greater risk of infection and
correspondingly, a higher ``success'' as an infectious agent. Some
scale-free networks present  positive assortativity. That
translates into the fact that highly connected nodes are connected
among them. This local structures can be used to model the
existence of core groups of high-risk individuals, that help to
maintain sexually transmitted diseases in a population dominated
by long-term monogamous relationships \cite{het}. Models of
disease spread through scale-free networks showed that the
infection is concentrated among the individuals with highest degree
\cite{pasto,new3}. One of the most surprising results is the one
found in Ref. \cite{pasto}. There, the authors show that no matter the
values taken by the relevant epidemiological parameter, there is
no epidemic threshold. Once installed in a scale-free network, the
disease will always propagate, independently of $R_0$.  Remember that when analyzed under the fully mixed
assumption, the studied SIS model has a threshold. The authors
perform analytic and numerical calculations of the propagation of
the disease, to show the lack of thresholds. Later, in Ref. \cite{lloy},
it was pointed out that networks with divergent second moments in
the degree distribution will show no epidemic threshold. The B A
network fulfills this condition. In Ref. \cite{lil1,lil2}, the authors
analyze the structure of different networks of sexual encounters,
to find that it has a pattern of contact closely related to a
power law. They also discuss the implications of such structure on
the propagation of venereal diseases

\begin{figure}[!h]
\begin{center}
\includegraphics[width=\columnwidth]{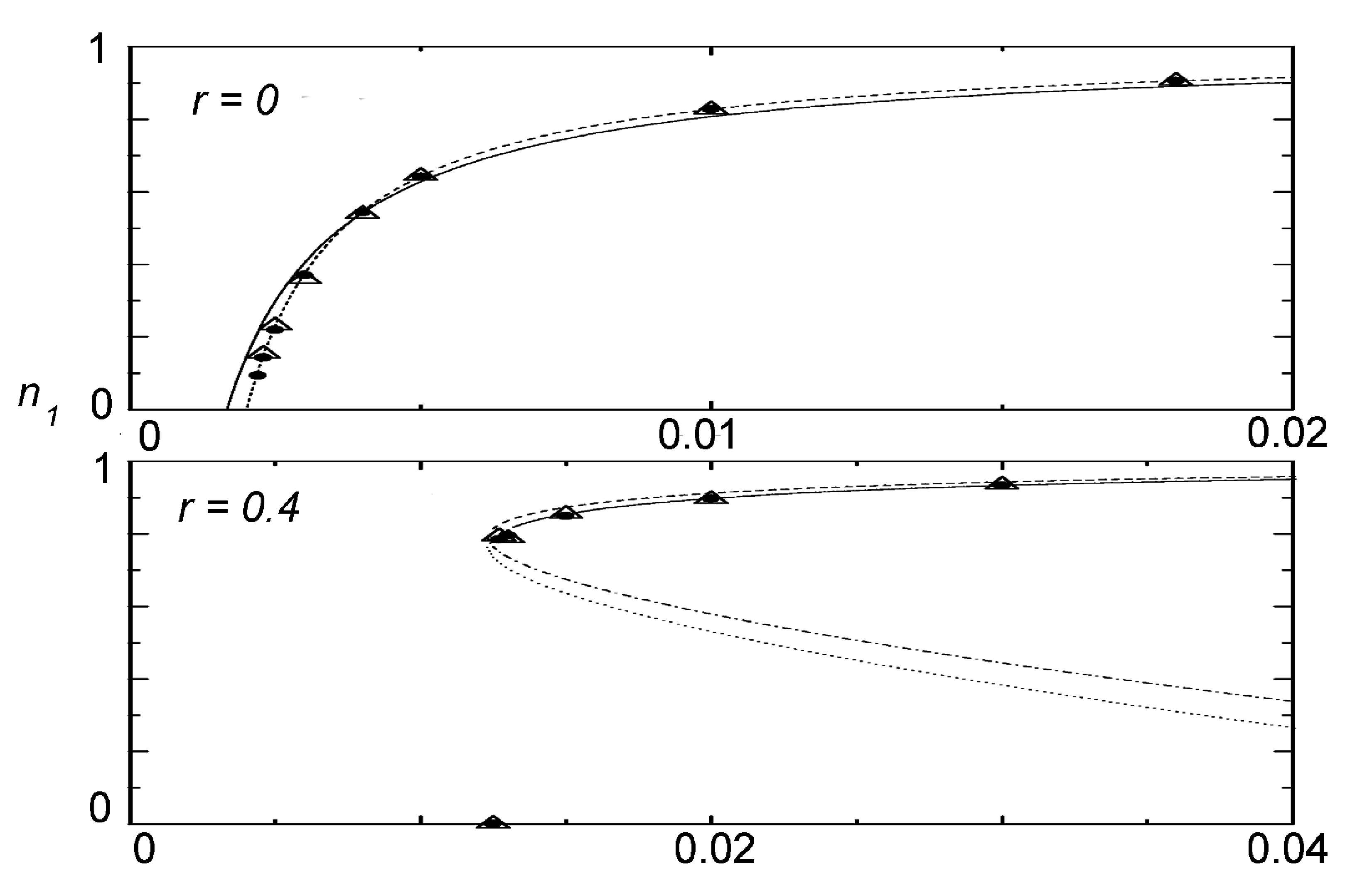}\\$\lambda$ 
\end{center}
 \caption{
These two  panels show the equilibrium fraction number of infected
individuals, as a function of the infectivity of the disease,
$\lambda$. Lines are analytic results, symbols are numerical
simulations. Adapted from Ref. \cite{zan2}.} \label{coev}
\end{figure}

{\bf Co-evolutionary networks.}
Co-evolutionary or adaptive networks take into account the own
dynamics of the social links. In some occasions, the
characteristic times associated to changes in social connections
 are comparable with the time scales of an epidemic process. Some
other times, the presence of n infectious core induces changes in
social links. Consider for example a case when the population
of susceptible individuals after learning about the  existence
of infectious individuals try to avoid them, or another case when
the health policies promote the isolation of infectious
individuals \cite{gros1}. The  behavior of models based on adaptive
network is determined by the interplay of two different dynamics
that  sometimes have competitive effects. On the one hand, we have
the dynamics of the disease propagation. On the other hand, the
network dynamics that operates to block the advance of the infection. The later is dominated by the rewiring rate of
the network, which affects the fraction of susceptible individuals connected to
infective ones. The most obvious choice is to eliminate the
infectious contacts of a susceptible individual by deleting or
replacing them with  noninfectious ones. The net effect is an
effective reduction of the infection rate. While static networks
typically predict either a single attracting endemic or
disease-free state, the adaptive networks show a new phenomenon, a
bistable situation shared by both states. The  bistability appears
for small rewiring rates \cite{gros1,shaw,zan1,zan2}. In Ref.
\cite{zan2}, the authors consider a contact switching dynamics. All
links connecting a susceptible agents with an infective one is
broken with a  rate $r$. The susceptible node is then connected to
a new  neighbor, randomly chosen among the entire population.
The authors show that reconnection can completely prevent an
epidemics, eliminating the disease. The main conclusion is that
the mechanism that they propose,  contact switching,  is a robust
and effective control strategy.
Figure \ref{coev} displays the results found in Ref. \cite{zan2}, where two completely different
 types of behavior can be distinguish as the rewiring parameter $r$ changes. The crossover from one regime to the other
is a second order phase transition.

\section{Immunization in networks}

Any epidemiological model can reproduce the fact that the number of individuals in a population
who are effectively immune to a given infection depends on the proportion
of previously infected individuals and the proportion who
have been efficiently vaccinated.

For some time, the  epidemiologists  knew about an emerging effect
called herd protection (or herd immunity). They discovered the occurrence of a global immunizing effect  verified  when the vaccination of a significant portion of a population provides protection for individuals who have not or cannot developed immunity.
Herd protection is particularly important for diseases transmitted from person to person. As the infection progresses through the social links,
its advance can be disrupted when many individuals are immune and their links to non immune subjects are no longer valid channels of propagation.  The net effect is that the greater the proportion of immune individuals is, the smaller the probability that a susceptible individual will come into contact with an infectious one.
The vaccinated individuals will not contract neither transmit the disease, thus establishing a firewall  between infected and susceptible individuals.

While taking profit from the  herd protection is far from being an optimal public health policy,  it is still taken into consideration when individuals cannot  be vaccinated due, for example, to  immune disorders or  allergies. The herd protection effect is equivalent to reduce the $R_0$ of a disease. There is a threshold value for the  proportion of necessary immune individuals in a population for the disease not to persist or propagate.
Its value depends on the  efficacy of the vaccine but also on the virulence of the disease and the contact rate.
If the herd effect reduces the risk of infection among the uninfected
enough, then the infection may no longer be sustainable within the
population and the infection may be eliminated.
In a real population, the emergence of herd immunity is closely related to the social architecture.
While many fully mixed models can describe the phenomenon,  the real effect is much more accurately reproduced by models based on Social Networks. One of the most expected result is to quantify how the shape of a social network can affect the level of vaccination required for herd immunity.
There is a related phenomenon, not discussed here, that consists in the propagation of real immunity from a vaccinated individual to a non vaccinated one.
This is called contact immunity and has been verified for several vaccines, such as the OPV \cite{opv}.

The models to quantify the success of immunization of the population propose a targeted immunization of the populations.

It is well established that immunization of randomly
selected individuals requires immunizing a very large fraction
 of the population, in order to arrest epidemics
that spread upon contact between infected individuals.

\begin{figure}[!h]
\centerline{\includegraphics[width=\columnwidth]{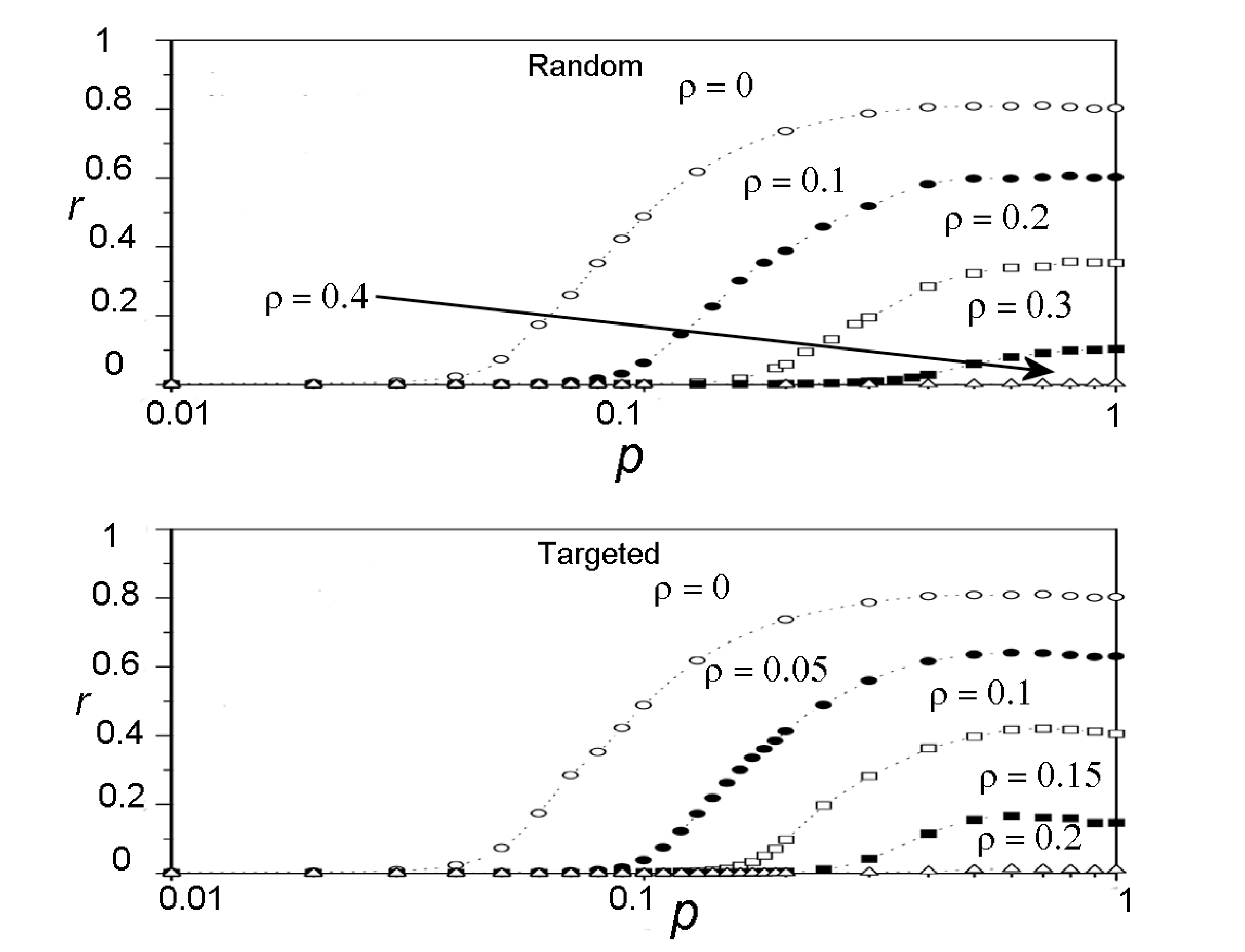}} \caption{
Fraction $r$ of the non-vaccinated population that becomes
infected during the disease propagation, as a function of the
disorder parameter $p$, for various levels of random immunization
(upper) and targeted immunization (bottom). Adapted from
Ref. \cite{kupim}. } \label{targ}
\end{figure}

In Ref. \cite{kupim}, the authors studied the effects of immunization on
an SIR epidemiological model evolving on a SW network. In the
absence of immunization, the model exhibits a transition from a
regime where the disease remains localized to a regime where it
spreads over a portion of the system. The effect of  immunization
reveals through two different phenomena. First, there is an overall
decrease in the fraction of the population affected by the
disease. Second, there is a shift of the transition point towards
higher values of the disorder. This can be easily understood as
the effective average number of susceptible  neighbors per
individual decreases.  Targeted immunization that is applied by
vaccinating  those individuals with the highest degree, produces a
substantial improvement in disease control. It is interesting to
point out that this improvement occurs even when the degree
distribution over small-world networks is relatively uniform, so
that the best connected sites do not monopolize a
disproportionately high number of links. Figure \ref{targ} shows an
example of the results found in Ref. \cite{kupim}, where the author
compare the amount of non vaccinated individuals that are infected
for different levels of vaccination, $\rho$, and different degrees
of disorder of the SW network $p$, as defined in Ref. \cite{ws}.

In a scale-free network, the existence of individuals of an
arbitrarily large degree implies that there is no level of uniform
random vaccination that can prevent an epidemic propagation, even
extremely high densities of randomly immunized individuals can
prevent a major epidemic outbreak. The discussed susceptibility of
these networks  to epidemic hinders the implementation of a
prevention strategy different from the trivial immunization of all
the population \cite{pasto,lloy,lloyim}.

Taking into account the inhomogeneous connectivity properties of
scale-free networks can help to develop successful immunization
strategies. The obvious choice is to  vaccinate individuals
according to their connectivity. A selective vaccination can be
very efficient, as targeting some of the super-spreaders can be
sufficient to prevent an epidemic \cite{pastim,lloy}.

The vaccination of  a small fraction
of these individuals increases quite dramatically the global tolerance to infections
of the network.

When comparing  the uniform and the targeted immunization
procedures \cite{pastim}, the results indicate that while uniform
immunization does not produce any observable  reduction of the
infection prevalence, the targeted immunization inhibits the
propagation of the infection  even at very low immunization
levels. These conclusions are particularly relevant when dealing
with sexually transmitted diseases, as  the number of sexual
partners of the individuals follows a distribution pattern close
to a power law.

Targeted immunization of the most highly connected individuals
\cite{albm,calla,pastim} proves to be effective, but requires
global information about the architecture of network that could
be  unavailable in many cases. In Ref. \cite{madar}, the authors
proposed a different immunization strategy that does not use
information about the degree of the nodes or other global
properties of the network but achieves the desired pattern of
immunization. The authors called it acquaintance immunization as
the targeted individuals are the acquaintances of randomly
selected nodes. The procedure  consists of choosing a random
fraction $p_i$ of the nodes,  selecting a random acquaintance per node
with whom they are in contact and  vaccinating them. The strategy
operates at the local level.  The fraction $p_i$ may be larger
than 1, for a node might be chosen more than once, but the
fraction of immunized nodes is always less than 1. This strategy
allows for a low vaccination level to achieve the immunization
threshold. The procedure is able to indirectly detect the most
connected individuals, as they are acquaintances of many nodes so
the probability  of being chosen for vaccination is higher.

\section{Final remarks}
The mathematical  modeling of the propagation of infectious
diseases transcends the academic interest. Any action pointing to
prevent a possible pandemic situation or to optimize the
vaccination strategies to achieve critical coverage are the core
of any public health policy. The understanding of the  behavior of
epidemics showed a sharp improvement during the last century,
boosted by the formulation of mathematical models. However, for a
long time, many important aspects regarding the epidemic processes
remained unexplained or out of the scope of the traditional
models. Perhaps, the most important one is the feedback mechanism
that develops between the social topology and the advance of an
infectious disease. The new  types of models developed during the
last decade made an important contribution to the field  by
incorporating a mean of describing the effect of the social
pattern. While a quantitative analysis of a real situation still
demands huge computational resources, the mathematical foundations
to develop it are already laid. The is too much to do yet, but the
breakthrough produced by these new models based on complex
networks is already undeniable.

\end{document}